\documentstyle{l-aa}
\input{epsf}
\begin{document}
\title{Ferromagnetism of dense matter and magnetic 
       properties of neutron stars }
\author{P. Haensel\inst{1,2}
\and S. Bonazzola\inst{2}}

\institute{N. Copernicus Astronomical Center, Polish
           Academy of Sciences, Bartycka 18, PL-00-716 Warszawa, Poland
\and
D{\'e}partement d'Astrophysique Relativiste et de Cosmologie, 
     UPR 176 du CNRS, Observatoire de Paris, 
Section de Meudon, 
 F-92195 Meudon Cedex, France}
\offprints{ P.Haensel}
\thesaurus{02.04.01, 08.14.1, 08.16.6, 02.07.01}
\date{received 5 March, 1996; accepted 22 April, 1996}
\maketitle
\markboth{P. Haensel and S. Bonazzola: 
          Ferromagnetism of dense matter and  neutron stars}{}
\begin{abstract}
Possible consequences of ferromagnetic
transition in dense matter 
suggested recently by Kutschera and W{\'o}jcik, 
for the magnetic properties of neutron stars, are studied. Specific model
of dense matter, in which a small admixture of protons is
completely polarized due to their interaction with neutrons, is
considered. 
 Magnetic field  of neutron stars
with a ferromagnetic core are calculated within 
the framework  of  general  relativity. 
 Two types of boundary conditions at the ferromagnetic core edge
are considered,  corresponding to normal and superconducting liquid
envelope, respectively. Numerical results for the neutron star
magnetic dipole moment are confronted with pulsar timing. To be
consistent with observations, ferromagnetic cores surrounded by
a non-superconducting envelope, should consist
of  weakly ordered ferromagnetic domains. If domains are
highly ordered, ferromagnetic core should be  screened by a
superconducting envelope. 

\end{abstract}
\section{Introduction }
Observations of radio and X -- ray pulsars indicate, that the
surface magnetic field of neutron stars can be as high as
$10^{12} - 10^{13}~$G. The strength of the internal magnetic
field of neutron stars is unknown; it could be much higher than
that of the surface one.

Standard present day models of magnetic field
identify its source with long--lived electric 
currents flowing in highly conductive neutron star matter. 
However, a complete scheme, based on such a 
model, which would explain both 
the origin and the evolution of neutron star magnetic field, is 
 still lacking (recent review on the present status  of this
topic can be found in Bhattacharya \& Srinivasan 1995). 

A very different type of model of neutron star magnetic field is
based on the hypothesis of existence of a ferromagnetic core in
the liquid interior of neutron star. First models of this type
were proposed just after the discovery of pulsars 
(Brownell \& Callaway 1969, Rice 1969, Silverstein 1969).
 Neutron star matter, approximated by a pure neutron matter, was
assumed to undergo a transition to a ferromagnetic state above
some critical density. Ferromagnetic neutron matter, carrying
high magnetization density, was there  a permanent source of a
superstrong magnetic field. This field, which would permeate stellar
interior, was  expected to extend outside the 
stellar surface as an observable
(mostly dipole) magnetic field of a pulsar. The qualitative
argument in favor of a ferromagnetic transition in 
sufficiently dense
neutron matter was based on Pauli exclusion principle, combined
with the repulsive character of the short range $n - n$
interaction. At high density, the  spin singlet ($S=0$) $n-n$
interaction becomes strongly repulsive, while the $S=1$ one, due
to Pauli principle, avoids the most repulsive, shortest range
contribution from the $l=0$ state (only odd $l$ states are
allowed in the $S=1$ channel). So, although in the ferromagnetic
transition the kinetic energy of the system increases, at
sufficiently high density this was believed to be 
 more than balanced by the
removal of the repulsive interactions in the $S=0$ states
(because of complete spin polarization, all
neutron pairs in a ferromagnetic phase 
are in the spin triplet state). The argument seemed to be
particularly convincing for the schematic $n-n$ interaction of
the infinite hard-core type (hard spheres gas). Some early
calculations, based on the schematic models of the $n-n$
interaction, seemed to show the existence of spontaneous spin
polarization in dense neutron matter above some critical density
(Brownell \& Callaway 1969, Rice 1969, Silverstein 1969, 
{\O}stgaard 1970; see, however, Clark \& Chao 1969, Clark 1969).  
However, further calculations, based on more
realistic $n-n$ interactions and/or more precise methods of
solution of the many body problem, ruled out possibility of
ferromagnetism of dense neutron matter 
(Pandharipande et al. 1972,  Haensel 1975, 
 Modarres \& Irvine 1979). 
  Such a
possibility was ruled out even in the most favorable case of the
hard-sphere model of neutron matter, where the contribution from
the  $l=1$ states was shown to be  very important at
superhigh densities (D{\c a}browski et al. 1978, 1979). 

It should be mentioned, that  in contrast to the realistic
calculations described above, the calculations done within 
the Hartree-Fock approximation using schematic {\it effective}
nucleon Hamiltionian (or Lagrangian) often showed a 
ferromagnetic transition in neutron matter at a few times
nuclear matter density. 
 Vidaurre et al. (1984) used non-relativistic effective $n-n$
interaction of the Skyrme type, and found a ferromagnetic transition
at about 1.5 times nuclear density. However, their result might
be just a consequence of the specific density dependence of the
spin dependent terms in the Skyrme interaction, which in most
cases leads to unphysical collapse (no lower bound for the
energy density) of dense polarized 
neutron matter (Kutschera \& W{\'o}jcik
 1994).  The spin stability of dense neutron matter  was also  
 studied within the relativistic Dirac-Hartree-Fock
approximation, with an 
effective  nucleon-meson Lagrangian, by Marcos et al (1991). These
authors find a ferromagnetic transition at several times nuclear
density, and point out that the presence of isovector mesons in
the mean-field Lagrangian is crucial for the possibility of
ferromagnetic transition. Let us notice, that the importance of
vector meson-coupling for the ocurrence of ferromagnetism was
pointed out some 20 years earlier (Kalman \& Lai 1971). 
In general, it seems that the ferromagnetic
transition in the mean-field theory models of dense neutron
matter is a consequence of a very specific density dependence of
the effective interactions characteristic of these models. 

In all studies reviewed above, neutron star matter was
approximated by pure neutron matter. The importance of the
presence of a small admixture of protons in neutron star matter
for the spin stability  was pointed out by Kutschera and
W{\'o}jcik (1989). They assumed that neutrons and protons in
neutron star  matter form normal Fermi liquids. 
 The ferromagnetic
state, corresponding to completely polarized protons and weakly
polarized neutrons, was shown to be energetically preferred over
the non--polarized one in two cases: for sufficiently strong
spin--spin component of the neutron--proton quasiparticle
interaction and/or for localized protons. Such conditions were
shown to be satisfied for numerous models of the neutron star
matter with low proton fraction. With their models of
ferromagnetic cores in neutron stars, Kutschera and W{\'o}jcik
were able to make specific predictions, concerning dependence
of the strength of the surface magnetic field on the neutron
star mass (Kutschera \& W{\'o}jcik 1992). The calculations
performed for strongly asymmetric nuclear matter within the
relativistic Dirac-Hartree-Fock approximation, using the mean
field nucleon-meson Lagrangian, confirmed that the presence of
an admixture of protons favors the ferromagnetic instability of
dense matter (Bernardos et al. 1995). 
   
At the densities exceeding the critical density for the onset of
ferromagnetism,  the ground state of neutron star matter
corresponds to a non-zero spin polarization. However, the
direction of the {\it local} spin polarization could vary within
the ferromagnetic core, which might have a domain
structure. The macroscopic magnetization, relevant for the
stellar magnetic field, may be  expected to result from some
 volume averaging, and should be consistent
with hydrostatic equilibrium of the star. We present in this
paper a self--consistent calculation of the magnetic  field of
neutron star, produced by a liquid, ferromagnetic core. Our
calculations, which are done within the framework of general relativity, 
 take into account constraints resulting from the
requirement of the hydrostatic equilibrium, as well as from 
realistic boundary conditions.

The paper is organized as follows. In Section 2 we present the
microscopic model of nucleon ferromagnetism. A model of
ferromagnetic neutron star core is described in Section 3. 
Physical conditions within the neutron star interior, relevant
for the properties of ferromagnetic neutron stars, are discussed
in Section 4. 
The calculation of  magnetic field of neutron star 
with a ferromagnetic core, performed in general relativity, is 
described in Section 5. 
Numerical results, obtained using specific assumptions
concerning physical conditions in neutron star interiors, are
presented in Section 6. 
 In Section 7 we confront our results with observations of radio
pulsars.  Our conclusions are presented in Section 8.
\section{Ferromagnetism of dense matter}
The possibility of ferromagnetism of dense nucleon matter results from
the existence of the spin--spin component of the quasiparticle
nucleon--nucleon interaction. We consider the simplest model of
neutron star matter, consisting of neutrons, protons and
electrons. At given baryon density, $n_{\rm b}$, the composition
of matter can be characterized by the proton fraction $x$, such
that proton density $n_{\rm p}=x n_{\rm b}=n_{\rm e}$, while
neutron density $n_{\rm n}=(1-x)n_{\rm b}$. The degree of spin
polarization of nucleons can be measured by the parameters
\begin{eqnarray}
s_{\rm n}&=&{n_{\rm n\uparrow}-n_{\rm n\downarrow}\over n_{\rm n}}~,
\nonumber\\
s_{\rm p}&=&{n_{\rm p\uparrow}-n_{\rm p\downarrow}\over n_{\rm p }}~,
\label{eq:Defin_s}
\end{eqnarray}
where $n_{\rm n\uparrow}$ is the number density of neutron
quasiparticles with $+1/2$ spin projection of the spin on the
spin quantization axis, etc. In the normal (i.e.,
non--ferromagnetic) state, the energy density reaches its
minimum at $s_{\rm n}=s_{\rm p}=0$. 

Our discussion of the properties of the spin--polarized 
neutron star matter 
will follow the paper of Kutschera and W{\'o}jcik (1989). 
 We assume that nucleons are normal (possible effects of nucleon
superfluidity  will be discussed in Sections 4, 6). 
 We
neglect the thermal effects, and calculate all thermodynamic
quantities at the $T=0$ approximation. The  change in the 
nucleon contribution to the energy density, 
implied by a small spin polarization of nucleons, $\delta E^N_{\rm
pol}$, can be calculated using the methods of the theory of
Fermi liquids,
\begin{eqnarray}
\delta E^N_{\rm pol}&=&
{n_{\rm n}^2\over 2N_n}(1+G_0^{\rm nn})s_{\rm n}^2+
{n_{\rm p}^2\over 2N_p}(1+G_0^{\rm pp})s_{\rm p}^2
\nonumber\\
& &~~~~~~ + n_{\rm n}n_{\rm p}g^{\rm np}_0 s_{\rm n}s_{\rm p}~,
\label{eq:Epol}
\end{eqnarray}
where $G_0^{\rm nn}$, $G_0^{\rm pp}$ and $g_0^{\rm np}$ are
Fermi liquid parameters, describing the spin--spin terms in the
qusiparticle nucleon--nucleon interaction in nucleon matter,
and $N_{\rm n}$, $N_{\rm p}$ are the densities of states at the
corresponding Fermi surfaces  for neutrons and protons,
respectively. While first and and second terms on the r.h.s. of
Eq.(\ref{eq:Epol}) are positive ($G_0^{\rm nn},~G_0^{\rm pp}>-1$), the
last term can be always made negative, irrespectively of the
sign of the $g_0^{\rm np}$ parameter. Notice, that the third
term, which results from the spin--spin interaction between
neutron and proton quasiparticles, vanishes for a
non--interacting system. 

If $\delta E^N_{\rm pol}<0$, the system is unstable with respect
to spin polarization. For small proton fraction ($x\ll 1$), the
approximate expression for $\delta E^N_{\rm pol}$ reads (Kutschera
\& W{\'o}jcik 1989, hereafter referred to as KW89)
\begin{equation}
\delta E^N_{\rm pol}\simeq 
{n_{\rm p}^2\over 2} 
\left(
{1\over N_{\rm p}} - 
{N_{\rm n}\over 1+G_0^{\rm nn}}(g_0^{\rm np})^2 
\right)
s_{\rm p}^2~. 
\label{eq:ENpol_app}
\end{equation}
Spontaneous spin polarization takes place for sufficiently
strong $g_0^{\rm np}$. The approximate condition for nucleon ferromagnetism  
reads thus
\begin{equation}
\left(g_0^{\rm np}\right)^2>
{1+G_0^{\rm nn}\over N_{\rm n}N_{\rm p}}~.
\label{eq:Ferro}
\end{equation}
If condition (\ref{eq:Ferro}) is satisfied, the ground state of matter
contains then completely polarized protons ($s_{\rm p}=1$), and
weakly polarized neutrons, 
\begin{equation}
s_{\rm n}=-x{g_0^{\rm np}N_{\rm n}\over 1 + G_0^{\rm nn}}~. 
\label{eq:snFerro}
\end{equation}
The values of the Fermi liquid parameters in strongly asymmetric
nuclear matter, as well as their density dependence, are very
uncertain. It seems to be reasonable to say, that their present
knowledge  does not exclude the situation, in which Equation
(\ref{eq:Ferro}) is satisfied above some critical density $n_*$. 

For sufficiently low proton fraction (such that protons can be
considered as impurities in neutron matter), the ground state of
nucleon matter could correspond to a non-uniform distribution of
neutrons, with proton localized (bound) in the neutron density
minima (Kutschera \& W{\'o}jcik 1990). 
 In the case of localized  protons, 
 the whole second term on the r.h.s. of Eq.(\ref{eq:Epol})
vanishes, and nucleon matter  is always unstable with respect to full
polarization of protons (KW89). 
Such a case corresponds most probably to the solid 
ferromagnetic core (Kutschera \& W{\'o}jcik 1995).

It should be stressed, that the results described above can be 
directly used in the neutron star calculations only under a
rather  unrealistic assumption of perfect space homogeneity of a
macroscopic element of matter. Only in such an idealized case
(which was actually assumed in  Kutschera \& W{\' o}jcik 1992) the
`microscopic' and `macroscopic' magnetizations will coincide. 
 It seems to be quite realistic to assume, that  
 the ferromagnetic phase of dense matter has some
 `domain structure'. In such a case, 
  microscopic results would be valid
 only within a single domain. The macroscopic values, relevant for the
stellar structure, would then correspond to an average over a sufficiently
large number of domains. To be more precise, the macroscopic
value of a quantity is then defined as an average over volume
element, which contains a sufficient number of domains, so that
the calculated value does not depend on the specific averaging
procedure. Of course, the element should be still small on the stellar
scale, so that the macroscopic value has a local character. 
 The macroscopic  quantities would have  to be consistent 
 with our assumption
about the axial symmetry of the star. 

It is well known, that ferromagnetism is an example of a 
second-order phase transition, and can be treated within the general
scheme of the theory of second-order phase transitions,
formulated by Landau (see, e.g., Section 39 of Landau et al.
1984). However, Landau theory of second-order phase transitions
asumes expansion of $E$ (or other suitable thermodynamic
potential) up to fourth order in polarization parameter.
Approximate expression for $\delta E^N_{\rm pol}$, Eq. (\ref{eq:Epol}), 
is
truncated at terms quadratic in polarization parameters. Such an
approximation is characteristic of the Fermi liquid theory, 
 in which only terms quadratic in deviations from the
normal, reference state, are conserved. Therefore, the model
proposed in KW89, if taken literally, cannot predict the
threshold behavior $s_{\rm p},\; s_{\rm n}\propto (n-n_*)^{1/2}$
for $n\longrightarrow n_*+0$, characteristic of a general case
of spontaneous polarization in ferromagnetic phase. 

The case of ferromagnetism with localized protons deserves a
special comment. The threshold density for proton localization,
$n_{\rm loc}$, corresponds to a phase transition from a liquid
mixture of protons, neutron and electrons to a proton crystal
immersed in a neutron and electron liquid. Such a liquid-solid
phase transitions is expected to be the first-order one (i.e.,
accompanied by the density and order parameter discontinuity).
However, for normal neutrons, localization implies complete
proton polarization, so that $n_*=n_{\rm loc}$. In such a case,
even for $n_{\rm b}\longrightarrow n_* +0$ protons remain fully
polarized. 
\section{Ferromagnetic neutron star cores}
For simplicity, we assume that
magnetization of the ferromagnetic core represents the only
source of stellar magnetic field (i.e., we neglect 
contribution to magnetic field resulting from electric
currents). 

 We  consider a simple model for the magnetization density of
polarized matter, ${\bf m}$, within the ferromagnetic core 
($n_{\rm b}>n_*$), as a
function of baryon density, $n_{\rm b}$. In  general, the
 spin contribution to the z-component of ${\bf m}$ 
 in the ferromagnetic phase is
given by 
\begin{equation}
\left(m_z\right)_{\rm spin}=
n_{\rm p}s_{\rm p}\mu_{\rm p} 
+n_{\rm n}s_{\rm n}\mu_{\rm n} ~,
\label{eq:mz_gen}
\end{equation}
where $\mu_{\rm p}$, $\mu_{\rm n}$ are intrinsic magnetic
moments of protons and neutrons, respectively, and we assumed
that z-axis is the spin quantization axis. We modeled
$\left(m_z\right)_{\rm spin}$ by a simple expression
\begin{equation}
\left(m_z\right)_{\rm spin}=\alpha\; 
 m_z^{\rm micr}~,
\label{eq:mz_alpha}
\end{equation}
where  $m_z^{\rm micr}$ corresponds to perfectly ordered
ferromagnetic state, considered  by Kutschera \& W{\'o}jcik
(1992, hereafter referred to as KW92), 
and dimensionless factor $\alpha\le 1$ describes the degree of
macroscopic polarization (degree of ordering of ferromagnetic
domains). 
  Our formula for ${\bf m}$, Eq.(\ref{eq:mz_alpha}), takes thus 
  into account a possibility
of  partial macroscopic ordering of nucleon spins
(partial ordering of ferromagnetic domains), including the case
of vanishing 
macroscopic  magnetization  within the ferromagnetic core
($\alpha=0$).   

Macroscopic magnetization density ${\bf m}$ is the source of
 magnetic field within  ferromagnetic neutron star cores. On the other
hand, the presence of non-vanishing ${\bf m}$ modifies the
energy density, due to coupling to magnetic field. Consistent
derivation of this coupling from microscopic theory shows, that
the additional term to be included is 
$-{\bf m}_{\rm spin}\cdot {\bf B}$ 
(de Groot \& Suttorp 1972, Chapter X; Carter 1982).
 If we take the $z$-axis along ${\bf B}$,
then minimization of energy requires $m_z B_z\geq 0 $. 
The consequences of this condition will be discussed in Section 6. 
\section{Physical conditions in neutron star interior}
In what follows, we consider the interior of neutron
star, of density $n_{\rm b}>n_{\rm h}=0.1~{\rm fm^{-3}}$. With
our assumptions, this  interior consists of a
ferromagnetic core of the density $n_{\rm b}>n_*$, and a
non--ferromagnetic liquid  envelope of the density 
$n_{\rm h}<n_{\rm b}<n_*$. 

Neutron stars are believed to be born as very hot objects, with
initial internal temperatures above $10^{11}$~K. Young neutron
star cools via neutrino emission. Initial temperature of
neutron star core can be expected to be  larger than, or of the
order of, the 
Curie temperature for the ferromagnetic transition, $T_{\rm
ferro}\sim {\rm~ few~ times}~10^{10}$~K (Haensel 1995, unpublished). 
Another important critical 
temperature is that for nucleon superfluidity, $T_{\rm sup}\sim
10^9$~K. It is thus reasonable to  expect, that the nucleon 
ferromagnetic transition occurs before superfluidity sets in, 
i.e., that ferromagnetism takes place in normal Fermi liquids.
Because $T_{\rm ferro}>T_{\rm sup}$, completely polarized proton
component cannot undergo singlet pairing at $T_{\rm sup}$. Also,
a fraction of polarized neutrons remains in the normal state,
due to their coupling to completely polarized protons. In such a
way, protons, and a small fraction (a few percent) of neutrons 
 may be expected to
be locked in a polarized state, due to energy barrier against the
depolarization of the ferromagnetic component.

It should be stressed, that in view of the uncertainties in the
many body theory of dense 
matter at supranuclear density, even the very existence of proton
superconductivity cannot be considered as absolutely certain. In
view of this, we may contemplate several possibilities. In the
first case, considered by KW92, ferromagnetic core is surrounded
by a liquid envelope of normal matter with 
$n_{\rm h}<n_{\rm b}<n_*$, 
 containing normal (i.e., non--superconducting) protons. 
 In the second case, liquid envelope is assumed to contain
superconducting protons. The superconductivity of the envelope
could strongly influence magnetic properties of
ferromagnetic neutron star (see Section 5, 6).
%
\section{Calculation of magnetic field}
 The calculation of the stellar magnetic field has been done
neglecting sources  other than magnetization of matter within
the ferromagnetic core. We took into account space curvature,
implied by the neutron star gravity. 

The first stage consisted in the calculation of the configuration of
hydrostatic equilibrium for a non-rotating neutron star with a
ferromagnetic core. In this calculation, we neglected the
influence of the presence of magnetic field on the equation of
state of the matter; this approximation will be justified
quantitatively in
Section 6. The  magnetic field vectors $H^{\rm j}$, $B^{\rm j}$
 were calculated by solving the equations of magnetostatics in curved
space. These equations are deduced from  the Maxwell equations
in the curved space in the static case (see, e.g., Eqs. (7.6) -
(7.7) of Carter (1980), or Thorne \& Macdonald (1982)),
\begin{equation}
\nabla_{\rm i} B^{\rm i}=0~,
\label{eq:divB}
\end{equation}
\begin{equation}
\epsilon^{\rm ijk}\nabla_{\rm j}(NH^{\rm k})=0~,
\label{eq:rotH}
\end{equation}
where $\nabla_{\rm i}$ denotes the covariant derivative on the
$t=const$ hyperspaces, $N$ is the lapse function, which in the
static case is given by $N=\sqrt{-g_{00}}$, and $\epsilon^{\rm
ijk}$ is the antisymmetric Levi-Civita tensor associated with
the 3-metric $\gamma_{\rm ij}$ on the $t=const$ hypersurfaces.  
Equation (\ref{eq:rotH})  will be satisfied, if and only if, $NH^{\rm i}$
is the gradient of some scalar function $\Psi$, so that
\begin{equation}
H^{\rm i}={1\over N}\nabla_{\rm i}\Psi~.
\label{eq:HPsi}
\end{equation}
Inserting (\ref{eq:HPsi}) into (\ref{eq:divB}) we get, 
using $B^{\rm i}=H^{\rm
i}+4\pi m^{\rm i}$, 
\begin{equation}
\widetilde{\Delta}\Psi=-4\pi\nabla_{\rm i}m^{\rm i}
+{1\over N}\nabla_{\rm i}N\nabla^{\rm i}\Psi~,
\label{eq:Lapl.Psi}
\end{equation}
where $\widetilde{\Delta}$ is the Laplacian operator in the
space curved by neutron star gravity, 
given by 
\begin{equation}
\widetilde\Delta{\Psi}=
{1\over \sqrt{\gamma}}\partial_{\rm ~i}\left(
\sqrt{\gamma}~\gamma^{\rm i j}\partial_{\rm ~j}\Psi
\right)~,
\label{eq:Lapl.def}
\end{equation}
where $\gamma\equiv {\rm det}(\gamma_{\rm ij})$.
 In view of the
smallness of the  deformation of the star, 
implied by the presence of magnetic field,
 we can consider the non-trivial elements of the 
metric  $\gamma_{\rm ij}$ as depending on the radial
coordinates only,
\begin{eqnarray}
\gamma_{11}&=&A^2(r)~,~~~~\gamma_{22}=r^2A^2(r)~,
\nonumber\\
\gamma_{33}&=&r^2\sin^2\theta~ A^2(r)~.
\label{eq:g_ij}
\end{eqnarray}
 Within this approximation, equation 
(\ref{eq:Lapl.def}) can be rewritten as
\begin{eqnarray}
\widetilde{\Delta}\Psi&=&
{1\over r^2 A^3}
{\partial\over\partial r}
\left(r^2 A{\partial\over \partial r}\Psi\right)
\nonumber\\
&+& 
{1\over A^2 r^2}\left({\partial^2\over\partial\theta^2}\Psi
+{\cos\theta\over\sin\theta}{\partial\over\partial\theta}\Psi\right)
~.
\label{eq:Lapl}
\end{eqnarray}
In what follows, we will make a simplifying assumption that 
 ${\bf m}(r)$ is parallel to
the symmetry axis of the star. The non-vanishing normal 
 components of magnetization in the local coordinate system 
 are then: 
$m_r=m\cos\theta$, $m_\theta=-m\sin\theta$ 
(where $m=\vert{\bf m}\vert$ depends on the $r$ coordinate only), 
so that 
$m_1=Am_r$, $m^1=m_r/A$, $m_2=Arm_\theta$, $m^2=m_\theta/(Ar)$. 
The divergence $\nabla_{\rm i}m^{\rm i}$ reads, in our
approximation, 
\begin{equation}
\nabla_{\rm i}m^{\rm i}=
{\cos\theta\over A^3}
{\partial\over\partial r}\left(A^2m\right)~.
\label{eq:div.m}
\end{equation}
The Eq. (\ref{eq:Lapl.Psi}) can thus be rewritten as
\begin{eqnarray}
& & {1\over r^2}
{\partial\over\partial r}
\left(r^2 A{\partial\over\partial r}\Psi\right)
+ {A\over r^2}\left(
{\partial^2\over\partial\theta^2}\Psi
+{\cos\theta\over\sin\theta}{\partial\over\partial\theta}\Psi
\right)
\nonumber\\
&=&-4\pi\cos\theta{\partial\over\partial r}
\left(A^2m\right)
 + {A\over N}{\partial N\over \partial r}
{\partial \Psi\over \partial r}~.
\label{eq:Eq1Psi}
\end{eqnarray}
Let us remind, that within our approximations $A$, $N$,
 and $m$ are
functions of $r$ only.  
Putting $\Psi=\Phi(r)\cos\theta$ we rewrite Eq. (\ref{eq:Eq1Psi}) 
in a more
suitable form:
\begin{eqnarray}
{1\over r^2}
\left[
{{\rm d}\over{\rm d} r}
\left(r^2A{{\rm d}\over{\rm d} r}\Phi\right) -2A\Phi
\right]
 &=& 
-4\pi{{\rm d}\over{\rm d} r}\left(A^2m\right)
\nonumber\\
&+&
{A\over N}{{\rm d}N\over {\rm d}r}{{\rm d}\Phi\over {\rm d}r}
~.
\label{eq:Eq2Psi}
\end{eqnarray}

As a consequence of general properties of the second--order
phase transitions, the
derivative of $m$ with respect to $r$ is singular at the edge of
ferromagnetic core, $r=r_*$. This is implied by the behavior
$m\longrightarrow const. (n_{\rm b}-n_*)^{1/2}$ for $n_{\rm
b}\longrightarrow n_* + 0$. To avoid
numerical problems stemming from this singularity, we rewrite
Eq. (\ref{eq:Eq2Psi}) as 
\begin{eqnarray}
{{\rm d}\over{\rm d} r}
\left[
Ar^2
{{\rm d}\over{\rm d} r}
\left(r^2\Phi\right)
\right]
&=&-4\pi{{\rm d}\over{\rm d} r}
\left(A^2m\right)
\nonumber\\
&+& {2\over r}\Phi{{\rm d}\over{\rm d} r}A
+{A\over N} {{\rm d}N\over {\rm d}r}
{{\rm d}\Phi\over {\rm d}r}~,
\label{eq:Eq3Psi}
\end{eqnarray}
which implies
\begin{eqnarray}
& & {A\over r^2}
{{\rm d}\over{\rm d} r}
\left(r^2A\Phi\right)
= -4\pi A^2m
\nonumber\\
&+& \int_0^r
 \left[
2{\Phi(r')\over r'}
{{\rm d}\over{\rm d} r'}A(r')
 + A(r'){{\rm d}N\over {\rm d}r'}
{{\rm d}\Phi\over {\rm d}r'}
\right]
{\rm d}r'~,
\label{eq:Eq4Psi}
\end{eqnarray}
so that
\begin{eqnarray}
\Phi(r)&=&
-{4\pi\over r^2}\int_0^r r'^2 A^2(r') m(r'){\rm d}r'
\nonumber\\
&+&{1\over r^2}
\int_0^r{{r'}^2\over A(r')} {\rm d}r'
\int_0^{r'}\times
\nonumber\\
& \times & \left[
{2\over r''}\Phi(r'')
{{\rm d}\over {\rm d}r''}A(r'')
  + A(r''){{\rm d}N\over {\rm d}r''}
{{\rm d}\Phi\over {\rm d}r''}
\right]
{\rm d}r''~.
\label{eq:Psi}
\end{eqnarray}

Let us consider first the case of normal envelope of the
ferromagnetic core. Then, 
 Equation (\ref{eq:Psi}) can be solved by iterations in all the space, 
$0\le r < \infty$, with regularity condition $\Phi(0)=0$, and
with boundary condition $\Phi(\infty)=0$. Solutions were
obtained using pseudospectral method, with three spatial grids,
in each of the three regions $0<r<r_*$, $r_*<r<R$ and
$R<r<\infty$, respectively. The method used for
compactification of the space  is the same as that used for the
calculation of the metric $g_{\rm ij}$ (see Bonazzola et al.
1993). Once $\Phi$ is known, the vectors $H^{\rm i}$ and $B^{\rm
i}$ are calculated from
\begin{eqnarray}
H_r(r,\theta) &=&{1\over AN}\Phi'(r)\cos\theta~,
\nonumber\\
H_\theta(r, \theta) &=&-{1\over ANr}\Phi(r)\sin\theta~,
\nonumber\\
B_r(r,\theta) &=& {1\over AN}\Phi'(r)\cos\theta+4\pi m_r~,
\nonumber\\
B_\theta(r,\theta) &=& -{1\over ANr}\Phi(r)\sin\theta
+4\pi m_\theta~.
\label{eq:formHB}
\end{eqnarray}
Magnetic permeability of normal neutron star matter is close to
unity;  for simplicity, we replaced it by one. 

In the second case, we assume that the ferromagnetic core 
is surrounded by liquid superconducting envelope
(with superconducting protons). Here, we can deal with two
distinct situations. If protons form a type-I superconductor, 
 then  ${\bf B}=0$ for  $n_{\rm h}<n_{\rm b}<n_*$ (Meissner
effect). This would correspond to a perfect screening of the
ferromagnetic core. It seems more likely, however, that the
magnetic field nucleated within the superconducting envelope as
an array of fluxoids, a situation characteristic for a type-II
superconductor (Baym et al. 1969, Sauls 1989). Such a
superconducting shell would behave as a diamagnetic envelope of
ferromagnetic core. 
%
\section{Magnetic field  under various physical conditions}
Our models of neutron star were calculated  assuming the BJI
model of the equation of state (EOS) of neutron star matter
(Malone et al. 1975). With this EOS, we obtain neutron star
models which are quite similar 
(for  $M\sim 1.4 - 1.8~M_\odot$) 
to those calculated using more recent EOS, such as the
UV14+TNI or FPR models considered by KW92. 
  For neutron star mass  $M=1.41~M_\odot$, we get the  radius  
  $R=12.11$ km, and central density 
  $n_{\rm centr}=0.58~{\rm fm^{-3}}$.
  
In our calculations of magnetic field, 
we assumed constant value  of the degree of
macroscopic polarization, $\alpha$, and a constant, small value
of the proton fraction, $x=0.05$. In view of linearity of Eq.
(\ref{eq:Eq2Psi}), the 
values of ${\bf B}$, ${\bf H}$ for a specific value of
$\alpha$ can be obtained from the $\alpha=1$ ones using 
${\bf B}=\alpha {\bf B}(\alpha=1)$, ${\bf H}=\alpha {\bf
H}(\alpha=1)$. For the model of KW92,
 the microscopic magnetization $m_z^{\rm micr}$ has a specific
dependence on the nucleon density, $n_{\rm b}=n_{\rm n}+n_{\rm
p}$.  We fitted the density  dependence  of $m_z^{\rm micr}$ 
by an analytic formula, which exhibited correct  threshold
 behavior for $n_{\rm b}\simeq n_*$. We assumed 
 $n_*=0.3~{\rm fm^{-3}}$. The specific density dependence 
 of $m_z^{\rm micr}$
 at intermediate densities results from the
density dependence of Fermi liquid parameters, combined with
opposite signs of $\mu_{\rm n}$ and $\mu_{\rm p}$, as well as
from the assumed dependence of the proton fraction on the
nucleon density. The quantity $m_z^{\rm micr}$, calculated in
KW92 
using Eq. (\ref{eq:mz_gen}), 
with $s_{\rm p}=1$ and $s_{\rm n}$ given by Eq. (\ref{eq:snFerro}),
changes sign at some $n_{\rm r}$. However, the condition 
${\bf m}{\bf B}\geq 0$ puts a definite condition on the sign of
$m_z$ on the stellar symmetry axis, which reads 
$m_z B_z\geq 0$. In view of this, vanishing of $m_z$ 
at some point (with $B_z\neq 0$) 
should not be accompanied by the change of sign of
$m_z$ in the vicinity of this point. In view of this, passing
through $n_{\rm r}$ does not change the sign of $m_z$, but
requires the reversal the signs of $s_{\rm n}$ and $s_{\rm p}$. 
  Let us notice, that the
nuclear contribution to the polarization energy, $\delta E^N_{\rm
pol}$, is invariant with respect to the change of the
orientation of spins, $s_{\rm n}\longrightarrow -s_{\rm n}$, 
$s_{\rm p}\longrightarrow -s_{\rm p}$, while such an inversion
of spins changes the sign of $m_z^{\rm micr}$. Therefore, 
 imposing the constraint $m_z^{\rm micr}B_z\geq 0$ in the 
 case of $\alpha=1$, when $m_z=m_z^{\rm micr}$,  has no effect on
the nuclear part of the energy density. However, it will have
important consequences for the magnetic properties of the
ferromagnetic core. 

Our basic calculations have been performed, assuming $\alpha=1$
and $M=1.41~M_\odot$. We considered the case
of a liquid envelope with normal protons.
 The radial distribution of $m_z$  is
displayed in Fig. 1.  In Fig. 2 
we show the the values of $H_z$ and $B_z$, respectively, on the
symmetry axis, as  functions of $z$. 

Let us notice, that relativistic effects in the ferromagnetic 
core are quite large. From equations (19) and (20), we obtain
relation between the values of $B_z$,$H_z$ and the lapse
function, $N$, at the center of the star,
\begin{equation}
{B_z(0)\over H_z(0)} = 1 - 3N(0)~.
\label{eq:HB_N}
\end{equation}
For our neutron star model we get 
 $B_z(0)/H_z(0)=-1.17$, to be compared with flat space-time
value $\left(B_z(0)/H_z(0)\right)_{\rm flat} = -2$ 
(M. Kutschera, private communication). 
The importance of relativistic effects results from
the fact, that the lapse function at the star center is
significantly lower than unity, $N(0)=0.618$.
\begin{figure}
\begin{center}
\leavevmode
\epsfxsize=8.0cm \epsfbox{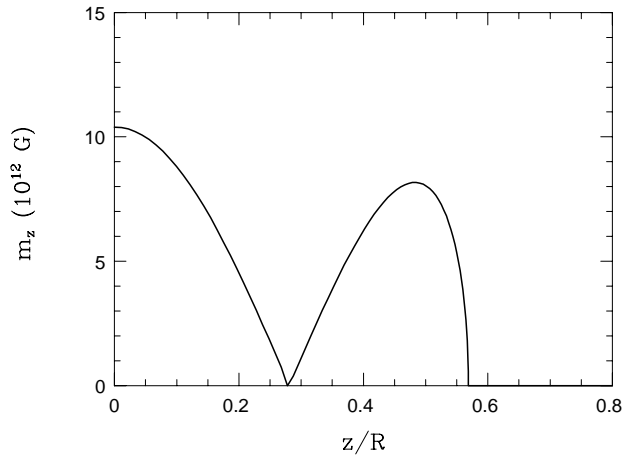}
\end{center}
\caption[]
{
 Magnetization of matter within the 
core  versus distance from the star center, for 
 the  $1.41~M_\odot$ model. Perfect ordering of ferromagnetic
domains ( $\alpha=1$ ) has been assumed.
}
\label{fig1}
\end{figure}
%

Let us notice, that
 the value of $B_z$ at magnetic poles ($z=\pm R$) 
 is significantly smaller ($\sim$ four times) 
 than its value at the ferromagnetic core edge.
The dipole magnetic moment of the star is 
${\cal M}^{\rm ferro}_{30}=15.1$ 
( here ${\cal M}_{30}$ is the
magnetic dipole moment in the units of $10^{30}~{\rm G~cm^3}$). 
At fixed
$M$, the values of $B_z({\rm pole})$ and ${\cal M}^{\rm ferro}$
for different values of the spin ordering parameter $\alpha$ can
be calculated  by multiplying by $\alpha$ the values obtained
for $\alpha=1$. 

Both $B_z({\rm pole})$ and ${\cal M}^{\rm ferro}$ depend on the
mass (central density) of the neutron star model. 
Obviously, we have 
${\cal M}^{\rm ferro}=0$ for 
$n_{\rm centr}<n_*$. However, with our choice of 
$n_*=0.3~{\rm fm^{-3}}$
and with our EOS, this implies that ferromagnetic core is
present in neutron stars  with $M>M_*=0.7~M_\odot$ - a condition
likely to be satisfied by radio pulsars. Both ${\cal M}^{\rm
ferro}$ and $B_z({\rm pole})$  increase with increasing stellar
mass. We have ${\cal M}_{30}=11.8$, $B_{z,12}({\rm pole})=11.7$
for $M=1.2~M_\odot$, and both quantities   icrease with
increasing mass up to 
${\cal M}_{30}=108$ and $B_{z,12}=174$ for the maximum allowable
mass, $M_{\rm max}=1.85~M_\odot$. Here, $B_{12}$ is the magnetic field
induction in the units of $10^{12}~$G. 
%
\begin{figure}
\begin{center}
\leavevmode
\epsfxsize=8cm \epsfbox{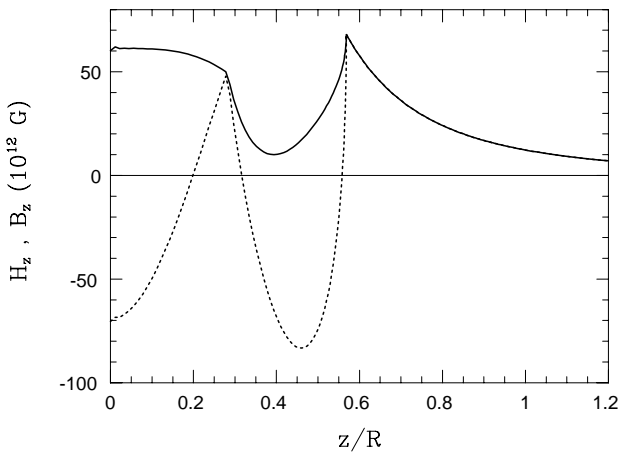}
\end{center}
\caption[]
{
 The values of $H_z$ (dotted line)  and $B_{z}$
(solid line) (in $10^{12}$~G) on  the z-axis, 
as a function of the distance from the star center, for the 
$1.41~M_\odot$ star. 
 Perfect ordering of ferromagnetic
domains ( $\alpha=1$ )  and normal liquid envelope 
 have been assumed. 
}
\label{fig2}
\end{figure}

It should be stressed, that in all cases  the deformation of 
 neutron star due to the presence of magnetic field is
indeed negligibly small. 
  For our model,
the maximum energy density of magnetic field created via the
ferromagnetic magnetization is comparable  to the gain in
the internal nuclear energy density resulting from the ferromagnetic
transition. A simple estimate, based on the formulae of Sect. 2,
 gives $\delta E^N_{\rm pol}\approx -3~10^{31}(n_{\rm
b}/n_0)^{2/3} (x/0.05)^{5/3}~{\rm erg/cm^3}$, which 
is very small compared to  the values of pressure near the neutron
star core, $P\sim 10^{35}~{\rm erg/cm^3}$.
%
\section{Confronting models with observations}
Results described in the  preceding section should be confronted
with existing information on the magnetic field of pulsars. When
confronting ferromagnetic neutron star models with observations, one 
should be aware of the fact, that even if some pulsars do
contain  ferromagnetic cores, such a core would be most probably
only one of the possible sources of pulsar magnetic field, the other one
being the long--living electric currents in the neutron star
interior.  

We will use a standard assumption,  that the slowing 
down of pulsar rotation is due to the
 loss of its rotational kinetic energy, implied by the emission
of the low frequency dipole radiation. We can then express the dipole
magnetic moment a pulsar of period $P$ and the time derivative
of the period, $\dot P$, as
\begin{equation}
{\cal M}_{\rm PSR}=\left(
{3c^3 I P {\dot P}\over 8\pi^2 \sin^2\beta}
\right)^{1\over 2}\;,
\label{eq:Mpsr}
\end{equation}
 where   $\beta$ is the angle between 
the magnetic and rotation
axis, and $I$ is the moment of inertia of neutron star. 
  Assuming canonical mass of radio pulsars, $1.4~M_\odot$, 
which for our EOS corresponds to $I=1.45\times 10^{45}~{\rm
g~cm^2}$,  
 we see that pulsar timing puts 
  limits on the magnetic moment of
ferromagnetic core (expressed here in the units of 
 $10^{30}~{\rm G~cm^3}$), 
\begin{equation}
{\cal M}^{\rm ferro}_{30}|\sin\beta|
<14~
\left({P\over 10~{\rm ms}}\right)^{1\over 2}\;
\left({{\dot P}\over 10^{-13}}\right)^{1\over 2}\;.
\label{eq:Mbound}
\end{equation}
Available timing data from Taylor et al. (1993) 
 yield  a very large range of  ${\cal M}_{30}|\sin\beta|$: 
 from   $2.3\times 10^{-3}$ for second most rapid 
 millisecond pulsar PSR B1957+20, to 
 $0.83$ for PSR B0154+61. 
  Excluding the case of
nearly perfect alignement of magnetic and rotation axes, we find
two situations consistent with  observations. In the first case,
ferromagnetic core is characterized by a low degree of
macroscopic ordering, which corresponds to 
$\alpha= 10^{-4} - 10^{-1}$. In the second 
case, the ferromagnetic core is
screened by the superconducting (type-I) envelope, and therefore
constraint  (\ref{eq:Mbound}) is irrelevant there. 
The case, when protons  in the liquid envelope form a type-II
superconductor would be an intermediate one; observational
constraint on ${\cal M}^{\rm ferro}$, Eq. (22), would then be
weakened due to partial screening of ferromagnetic core. 
\parindent 0pt
\vskip 1cm
{\bf 8. Discussion and conclusions}
\parindent 21pt
\par
\vskip 0.5cm
In the ferromagnetic model of Kutschera and W{\'o}jcik (1992),
consistency of the obtained values of surface magnetic field
with pulsar timing was obtained via cancellations of the
contributions to the stellar dipole magnetic moment coming from
the inner and outer regions of the ferromagnetic cores. 
In particular, 
the low values of $B$ for millisecond pulsars were explained by 
 assuming that the magnetic moment of their ferromagnetic cores
was some three orders of magnitude smaller than that of ordinary
pulsars. According to Kutschera and W{\'o}jcik (1992), such a
situation was made possible due to the cancellations, resulting
from  the change of sign within the ferromagnetic core.
 This was
possible only within a very narrow interval of neutron star
masses ($\Delta M\sim 0.1~M_\odot$). 

Consistent calculations of the magnetic field produced by the
ferromagnetic core of neutron star, performed in the present
paper,  show that the projection of
the macroscopic magnetization onto the symmetry axis cannot
change sign within the core. In view of this, consistency with
pulsar timing can be achieved only under some  specific conditions. If
the liquid envelope surrounding ferromagnetic core is
non-superconducting, than the ferromagnetic phase has to be
highly disordered, with the degree of macroscopic polarization
varying from $\sim 10^{-2}$ for young pulsars, to
$\sim 10^{-4}$ for the millisecond pulsars. In the case,
when the liquid envelope is a type-I superconductor, the
ferromagnetic core produces no external magnetic field, which
could be 
then produced exclusively by the currents within the neutron
star crust. 

Recently, Kotlorz \& Kutschera (1994) pointed out an interesting
possibility of existence of ferromagnetic pion-condensed quark
cores, within sufficiently massive neutron stars. Their results
imply ${\cal M}^{\rm ferro}_{30}({\rm quark})\sim 10^3$. Such huge
values of ${\cal M}^{\rm ferro}$ can be reconcilied with pulsar
timing only if either pion-condensed quark core has very low
degree of macroscopic spin ordering, $\alpha<10^{-4}$, or if
this core is screened by the superconducting nucleon envelope. 
\begin{acknowledgements}
 We express our deep gratitude to Eric Gourgoulhon for his
helpful  remarks  concerning Maxwell 
equations in the curved
space, and for the careful reading of the manuscript. 
We are also grateful to Marek Kutschera for pointing out the
importance of relativistic effects.  
This research has been supported in part by the Polish  State 
Committee for
Scientific Research (KBN) grant and by PICS/CNRS
198 `Astronomie Pologne'.
\end{acknowledgements}


\begin{thebibliography}{}
\bibitem[1969]{bpp69}
Baym G., Pethick C., Pines D., 1969, Nature 224, 673
\bibitem[]{}
Behera, B., Satpathy, R.K., 1979, J. Phys. G 5, 1085
\bibitem[]{}
Bernardos, P., Marcos, S., Niembro, R., Quelle, M.L., 1995, 
Phys. Lett. B 356, 175
\bibitem[]{}
Bhattacharya, D., Srinivasan, G., 1995, 
in: W.H. Lewin, J. van Paradijs, 
E.P.J. van den Heuvel (eds) X-Ray Binaries. 
Cambridge UP, Cambridge, p. 495
\bibitem[]{}
Brownell, B.H., Callaway, J., 1969, Nuovo Cimento 60B, 169
\bibitem[]{}
Bonazzola, S., Gourgoulhon, E., Salgado, M., Marck, J.A., 1993,
A\& A 278, 421 
\bibitem[]{}
Carter, B., 1980, Proc. R. Soc. Lond. A372, 169
\bibitem[]{}
Clark, J.W., 1969, Phys. Rev. Lett. 23, 1463
\bibitem[]{}
Clark, J.W., Chao, N.-C., 1969, Lettere Nuovo Cimento 2, 185
\bibitem[]{}
D{\c a}browski, J., Piechocki, W., Ro{\.z}ynek, J., Haensel, P., 
1978,  Phys. Rev. C17, 1516
\bibitem[]{}
D{\c a}browski, J., Piechocki, W., Ro{\.z}ynek, J., Haensel, P., 
1979,  Acta Phys. Polonica B10, 83
\bibitem[]{}
de Groot, S.R., Suttorp, L.G., 1972, 
Foundations of Electrodynamics. Amsterdam: North Holland
\bibitem[]{}
Haensel, P., 1975, Phys. Rev. C11, 1822
\bibitem[]{}
Kalman,G., Lai, S.T., 1971, Phys. Lett. A 34, 75
\bibitem[]{}
Kotlorz, A., Kutschera, M., 1994, Acta. Phys. Polonica B 25, 859
\bibitem[]{}
Kutschera, M., W{\'o}jcik, W., 1989, Phys. Lett. B 223, 11
\bibitem[]{}
Kutschera, M., W{\'o}jcik, W., 1992, Acta Phys. Polonica B 23, 947
\bibitem[]{}
Kutschera, M., W{\'o}jcik, W., 1995, Nucl. Phys. A 581, 706
\bibitem[]{}
Malone, R.C., Johnson, M.B., Bethe, H.A., 1975, ApJ 199, 741
\bibitem[]{}
Marcos, S., Niembro, R., Quelle, M.L., Navarro, J., 1991, 
Phys. Lett. B 271, 277
\bibitem[]{}
Modarres, M., Irvine, J.M., 1979, J.Phys.G 5, L133
\bibitem[]{}
{\O}stgaard, E., 1970, Nucl. Phys. A154, 202
\bibitem[]{}
Pandharipande, V.R., Garde, V.K., Srivastava, J.K., 1972, 
Phys. Lett. 38B, 485
\bibitem[]{}
Press, W.H., Thorne, K.S., 1972, ARA\&A 10, 335
\bibitem[]{}
Rice, M.J., 1969, Phys. Lett. 29A, 637
\bibitem[]{}
Sauls, J.A., 1989, in: 
H. {\"Ogelman} and E.P.J. van den Heuvel (eds) 
Timing Neutron Stars. Dordrecht: 
 Kluwer Academic Publishers, p. 457
\bibitem[]{}
Shapiro, S.L., Teukolsky, S.A., 1983, Black Holes, White Dwarfs
and Neutron Stars. New York:  Wiley and Sons
\bibitem[]{}
Vidaurre, A., Navarro, J., Bernab{\'e}u, J., 1984, A\&A 135, 361 
\bibitem[]{}
Taylor, J.H., Manchester, R.N., Lyne, A.G., 
1993, ApJ Suppl. 88, 829
\bibitem[]{}
Thorne, K.S., 1980, Rev. Mod. Phys. 52, 285
\bibitem[]{}
Thorne, K.S., Macdonald, D., 1982, MNRAS 198, 339
\bibitem[]{}
\end{thebibliography}
\end{document}